\documentclass[epsfig]{mn2e}
\usepackage{times,graphics,epsfig}
\newif\ifAMStwofonts

\title{Coupling between QPOs and broadband noise components in GRS
1915+105}

\author[Maccarone et al.] {Thomas J. Maccarone, Philip Uttley\\
University of Southampton, School of Physics and Astronomy, Highfield
Campus, Southampton, Hampshire, SO17 1BJ \newauthor Michiel van der
Klis, Rudy A. D. Wijnands\\ Astronomical Institute ``Anton Pannekoek'',
University of Amsterdam, Kruislaan 403, 1098 SJ, Amsterdam, The
Netherlands \newauthor Paolo S. Coppi\\Yale University,Department of
Astronomy,PO Box 208101, New Haven CT 06520-8101}

\date{}

\begin{document}

\maketitle

\label{firstpage}

\def\simlt{\mathrel{\rlap{\lower 3pt\hbox{$\sim$}}
        \raise 2.0pt\hbox{$<$}}}
\def\simgt{\mathrel{\rlap{\lower 3pt\hbox{$\sim$}}
        \raise 2.0pt\hbox{$>$}}}

\input epsf

\begin{abstract}

We explore the use of the bispectrum for understanding quasiperiodic
oscillations.  The bispectrum is a statistic which probes the
relations between the relative phases of the Fourier spectrum at
different frequencies.  The use of the bispectrum allows us to break
the degeneracies between different models for time series which
produce identical power spectra.  We look at data from several
observations of GRS~1915+105 when the source shows strong
quasi-periodic oscillations and strong broadband noise components in
its power spectrum.  We show that, despite strong similarities in the
power spectrum, the bispectra can differ strongly.  In all cases,
there are frequency ranges where the bicoherence, a measure of
nonlinearity, is strong for frequencies involving the frequency of the
quasi-periodic oscillations, indicating that the quasi-periodic
oscillations are coupled to the noise components, rather than being
generated independently.  We compare the bicoherences from the data to
simple models, finding some qualitative similarities.

\end{abstract}

\section{Introduction}

Recent studies of the variability properties of accreting black holes
and neutron stars have shown that these systems' Fourier power spectra
in certain spectral states are well-described by a sum of many
Lorentzian components (see e.g. Olive et al. 1998; Nowak 2000;
Belloni, Psaltis \& van der Klis 2002; van Straaten et al. 2002;
Pottschmidt et al. 2003).  Essentially the same components seem to
appear in nearly all sources, and at nearly all luminosities, and
their frequencies tend to be well correlated with one another and with
the source spectral states (Wijnands \& van der Klis 1999; Psaltis,
Belloni \& van der Klis 1999).  This phenomenology suggests that there
may be a single origin for most of the variability features in the
power spectra of accreting black holes and neutron stars.  Similar
correlations in frequencies between the different components seem to
apply even to accreting white dwarfs (e.g. Mauche 2002; Warner, Woudt
\& Pretorius 2003).

Coupling between different components in the power spectrum has been
suggested in many theoretical contexts.  In a shot noise model
(e.g. Terrell 1972), variability components on the timescale of the
shot will be correlated with one another, resulting in, e.g. fast rise
exponential decay or exponential rise, fast decay profiles, but there
should be no coupling on timescales longer than those of the shots.
More sophisticated models of variability, for example, self-organized
criticality (e.g. Takeuchi, Mineshige \& Negoro 1995; Takeuchi \&
Mineshige 1997), predict correlations between the arrival times and/or
intensities of the shots.  Other related ``reservoir'' models
(e.g. Merloni \& Fabian 2001; Maccarone \& Coppi 2002; Malzac, Merloni
\& Fabian 2004), where the accretion disk and/or the relativistic jet
taps an energy supply effectively enough to reduce the available
energy for future emission also predict variability correlated over
many frequencies.  Resonance models for producing quasi-periodic
oscillations should also clearly produce non-linear variability
(e.g. Psaltis \& Norman 2001; Abramowicz \& Kluzniak 2001; Schnittman
\& Bertschinger 2003; Maccarone \& Schnittman 2005), while
non-resonant models for producing the same QPOs (e.g. Chen \& Taam
1992,1995; Rezzolla et al. 2003; Giannios \& Spruit 2004) could, but
need not show coupling between the different frequencies.  Propagation
models, where disturbances move through an accretion disc, also should
produce non-linear coupling of variability components, as the
variability properties are modified at each annulus (Lyubarskii 1997).

A key to verifying and understanding this possible unified origin for
variability is to go beyond the simple power spectrum and begin
studying their non-linear variability.  The first attempts at this
failed to detect signatures of non-linearity, with the methods
including the time skewness (Priedhorsky et al. 1978) and searches for
a low dimensional chaotic attractor (Lochner, Swank \& Szymkowiak
1989).  On the other hand, more recent work with better light curves
did establish that the light curves of Cygnus X-1 are not time
reversible (Timmer et al. 2000; Maccarone \& Coppi 2002), that there
exists a correlation between rms amplitude and flux of a source that
is inconsistent with pure shot noise models (Uttley \& McHardy 2001),
that there is coupling between variability components on all
observable time scales (Maccarone \& Coppi 2002), and that in some
cases, there is a low dimensional chaotic attractor after all (Misra
et al. 2004), and the observed light curves may be described by a
Lorenz system (Misra et al. 2006).  With this in mind, we now approach
looking at the properties of the coupling between quasi-periodic
components and noise components in GRS~1915+105, a bright Galactic
X-ray binary with an accreting black hole which has been the subject
of several long observations with the {\it Rossi X-ray Timing Explorer
\rm(\it{RXTE}\rm)}.  

We treat this work as a pilot study -- the first attempt to apply the
bispectrum to astronomical data with strong quasi-periodic
oscillations.  As such, we aim to illustrate the power of the
technique by showing data and simple models that can give very similar
power spectra, and very different bicoherences, but consider it beyond
the scope of the work to try to fit models to the data precisely.  We
present computations of the bicoherence for several observations of
GRS~1915+105, discuss the meaning of the bicoherence, and present a
few toy models for the bicoherence which we compare with the data.

\section{Data Used}
For this pilot study, we present the results from three representative
observations of GRS~1915+105; our goals in this paper are not to make
a complete characterisation of the bicoherence properties of all
sources in all states, but rather to demonstrate of the utility of the
technique for identifying phenomenological differences between
apparently rather similar lightcurves with rather similar power
spectra, and to show a few theoretical models that produce
bicoherences similar to some of the observations.  All these data are
from variability class $\chi$ in the classification scheme of Belloni
et al. (2000).  In this class, GRS~1915+105 stays steadily in state C,
its ``low/hard'' state -- the ``hard very high state'' in the
nomenclature of Fender \& Belloni (2004), or the hard intermediate
state in the more recent classification of Homan \& Belloni (2005). In
this state, GRS~1915+105 shows a strong power law component in its
X-ray spectrum, and high rms amplitude quasi-periodic oscillations at
about 0.5-10 Hz, often with powerful harmonics (see Fender \& Belloni
2004 and references within).  The broadband noise component in the
power spectrum containts a greater fraction of the rms amplitude of
variability than do the QPOs and harmonics.

The RXTE observation identification numbers of the three observations
presented are 10408-01-25-00 (taken 19 July 1996), 20402-01-15-00
(taken 9 Feb 1997), and 30184-01-01-000 (taken 4 April 1998).
Standard screening has been applied to the data to ensure no usage of
data taken during Earth occulations or periods of high offset.  The
data used are the single bit mode data with the lowest set of energy
channels: channels 0-35 are used for 10408-01-25-00 and
30184-01-01-000, while channels 0-13 are used for 20402-01-15-00.
These correspond approximately to the energy range from 2-13 keV and
2-5 keV, respectively.  We have done checks of the other data modes
with different energy channel ranges and have found no qualitative
differences and only minor quantitative differences in the properties
of the bicoherence as a function of photon energy, so we have used the
lowest energy set of channels in a single data mode for each
observation for convenience -- we will not attempt to make
interpretations of the data at a level of detail approaching the
differences caused by changing energy band used.  The data are then
Fourier transformed with 4096 elements per transform, and time
resolution of 1/64 seconds, 1/128 and 1/512 seconds, respectively.
These are chosen so that the quasi-periodic oscillation is near the
middle of the range of Fourier bins in each observation (i.e. a lower
time resolution is used when the QPO has a lower frequency).  The
integration times used are 8640, 10240, and 8160 seconds,
respectively, with only the first part of observation 30184-01-01-000
used.  This decision was originally made due to memory limitations
when the calculations were first made, and the decision not to re-do
the calculation was made because there is significant frequency drift
over the full observation, and the effects of frequency drift are
ameliorated by using shorter total integrations.  The Fourier
transforms are then combined to produce the bispectra and bicoherences
of these data sets.

We have examined the amount of frequency drift in the different
observations.  For observation 10408-01-25-00 and 20402-01-15-00, the
peak frequency in the power spectrum varies by less than 10\% over the
full integrations -- less than the widths of the QPOs.  For
30184-01-01-000, there is substantial variation of the peak frequency
if we consider the full observation -- with peak frequencies ranging
from about 2.7 Hz to 4 Hz.  However, in the part of the data set we
consider, the variation is only from about 3.3 Hz to 4 Hz.  Some minor
effects from this variation can be seen in the computed bicoherences,
and we discuss them below.

\section{Statistical Methods}
For this work, we will focus on computing the bispectrum and the
closely related bicoherence.  The bispectrum computed from a time
series broken into $K$ segments of equal length is defined as:
\begin{equation}
B(k,l)=\frac{1}{K} \sum_{i=0}^{K-1} X_i(k)X_i(l)X^*_i(k+l),
\end{equation}
where $X_i(k)$ is the frequency $k$ component of the discrete Fourier
transform of the $i$th time series (e.g. Mendel 1991; Fackrell 1996
and references within).  The bispectrum is well defined only for $k+l$
less than or equal to the Nyquist frequency of the data used to
compute it, and is defined nontrivially only for $k<l$, since
$B(k,l)=B(l,k)$ -- although in this paper, we will plot the
bicoherence for $k>l$ and for $l>k$ to help guide the eye to features
in the plots.  The expectation value of the bispectrum is unaffected
by additive Gaussian noise, although its variance will increase for a
noisy signal.  Poisson noise can affect the bicoherence for
frequencies where the Poisson level is high compared with the level of
intrinsic variability -- the amplitude of the Poisson noise correlates
with the count rate, so that there will be phase coupling between the
instrinsic variability of the source's signal, and the noise level in
the Poisson component (see e.g. Appendix C of Uttley et al. 2005).  In
this paper, we focus on low frequency variability of a highly variable
source, so Poisson noise is unimportant.

The definition of the bispectrum gives its absolute phase a
well-defined meaning, in constrast to the phase of the Fourier
spectrum which has a dependence on an arbitary reference time.  One
can then attempt to determine the degree of constancy of that phase,
making use of a related quantity, the bicoherence -- the magnitude of
the bispectrum, normalised to lie between 0 and 1.  Defined
analogously to the cross-coherence function (e.g. Nowak \& Vaughan
1996), it is the vector sum of a series of bispectrum measurements
divided by the sum of the magnitudes of the individual measurements.
If the biphase (the phase of the bispectrum) remains constant over
time, then the bicoherence will have a value of unity, while if the
phase is random, then the bicoherence will approach zero in the limit
of an infinite number of measurements.  Specifically, the squared
bicoherence, $b^2$ is defined as:
\begin{equation}
b^2(k,l) = \frac{\left|\sum{X_i(k)X_i(l)X^*_i(k+l)}\right|^2}{\sum{\left|X_i(k)X_i(l)\right|^2}\sum{\left|X_i(k+l)\right|^2}}.
\end{equation}

This normalization of the bicoherence was proposed by Kim \& Powers
(1979).  If it shows variation as a function of the frequencies used,
the time series analysed is nonlinear.  It has been shown that other
normalisations are more likely to detect nonlinearity under certain
specific circumstances (Hinich \& Wolinsky 2004), but we use the Kim
\& Powers (1979) normalisation here to retain consistency with past
work.  In previous work where we were studying variability components
which were very broad in the power spectrum (Maccarone \& Coppi 2002),
we binned the data to make a one dimensional function of $k+l$, and
substantially re-binned the bispectrum values over frequencies and
then compared the results with model predictions.  Since that work, we
have become aware of a correction which is, in principle, important
for studying aperiodic variability with the bicoherence.  The maximum
value of the bicoherence is suppressed by smearing of many frequencies
into a single bin in the discrete Fourier transform (S. Vaughan,
private communication).  This suppression cannot be calculated in a
straightforward way (see e.g. Greb \& Rusbridge 1988).  We note also,
though, that since comparisons in Maccarone \& Coppi (2002) were made
only with model calculations made with the same time binning as the
real data, these effects, whatever they may be, are the same for the
real data and the simulated data, and hence the conclusions of that
paper are not affected substantially.  Therefore, to make quantitative
comparisons of real data with simulated bicoherences, it is important
to use the same frequency binning in both cases.

A non-zero bicoherence can be used to rule out, for example, models
where the variability on different timescales comes from independent
Gaussian model components which are then added linearly.  Beyond that,
it is often difficult to interpret the results of bicoherence
analysis, but comparison of observed results with results from
simulated light curves can be a very powerful tool.  As shown already,
for example, in Maccarone \& Schnittman (2005), light curves with very
similar power spectra can have easily distinguishable bicoherence
plots.

For this work, we will consider variability only at frequencies which
are low enough that the intrinsic variability of the source is strong,
so that neither Poisson noise nor dead time affects the Fourier
spectrum substantially.  We do note that these problems may be more
important at higher frequencies and that they will need to be
considered in future analyses.

\section{Bicoherence results}
\begin{figure}
\begin{center}
\renewcommand{\epsfsize}[2]{0.7#1}
\epsfig{file=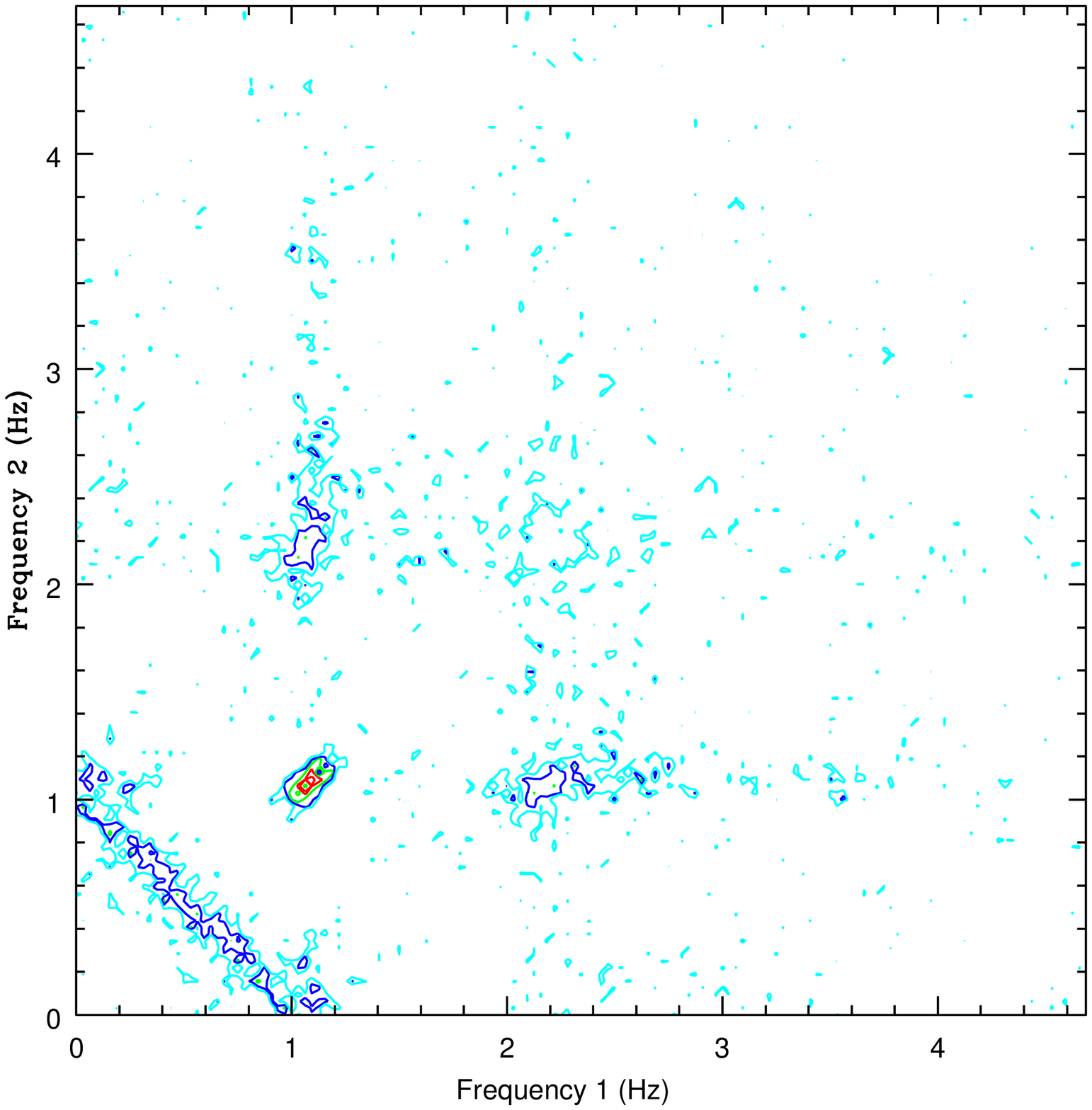,width=6cm}
\epsfig{file=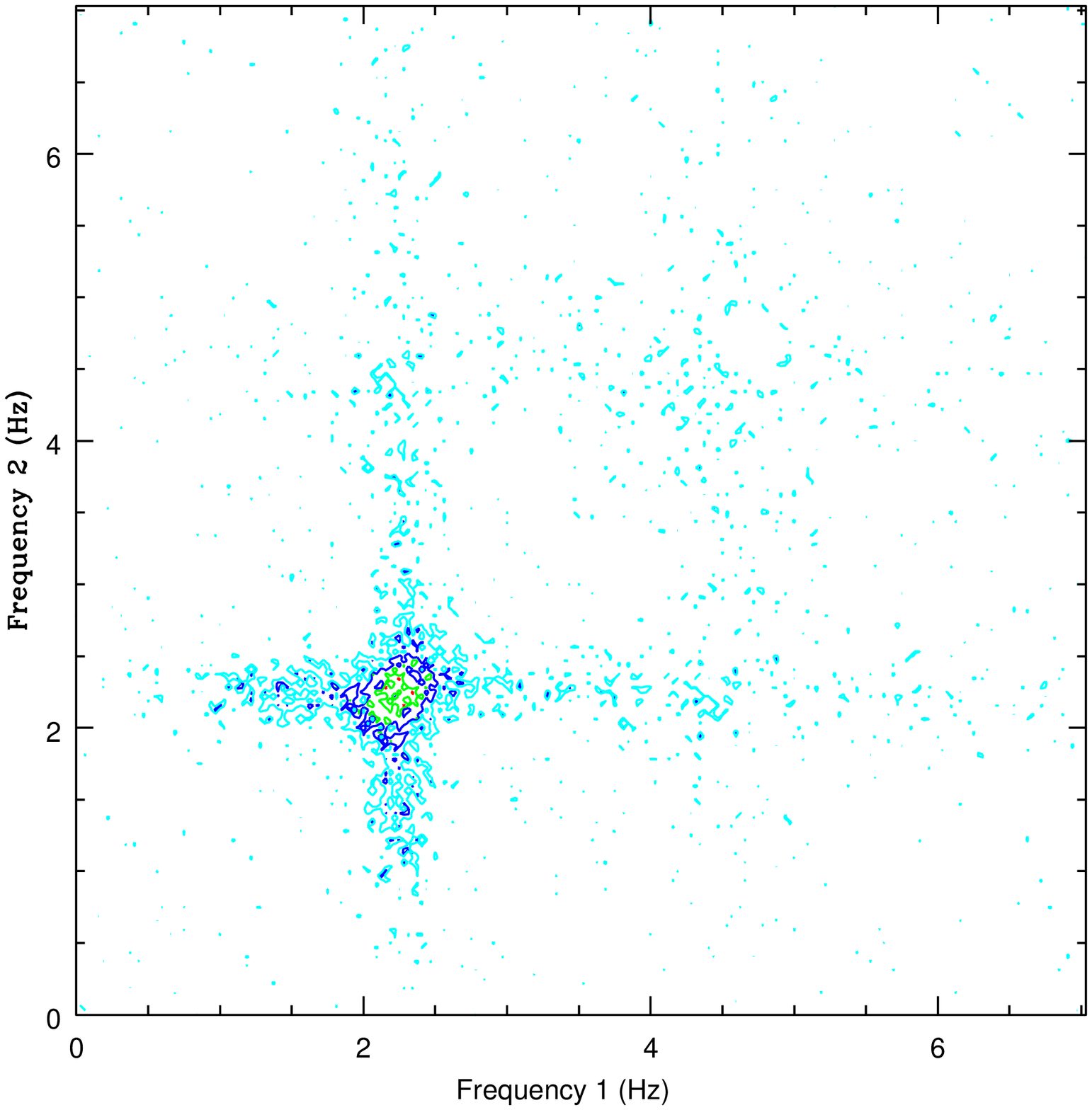,width=6cm}
\epsfig{file=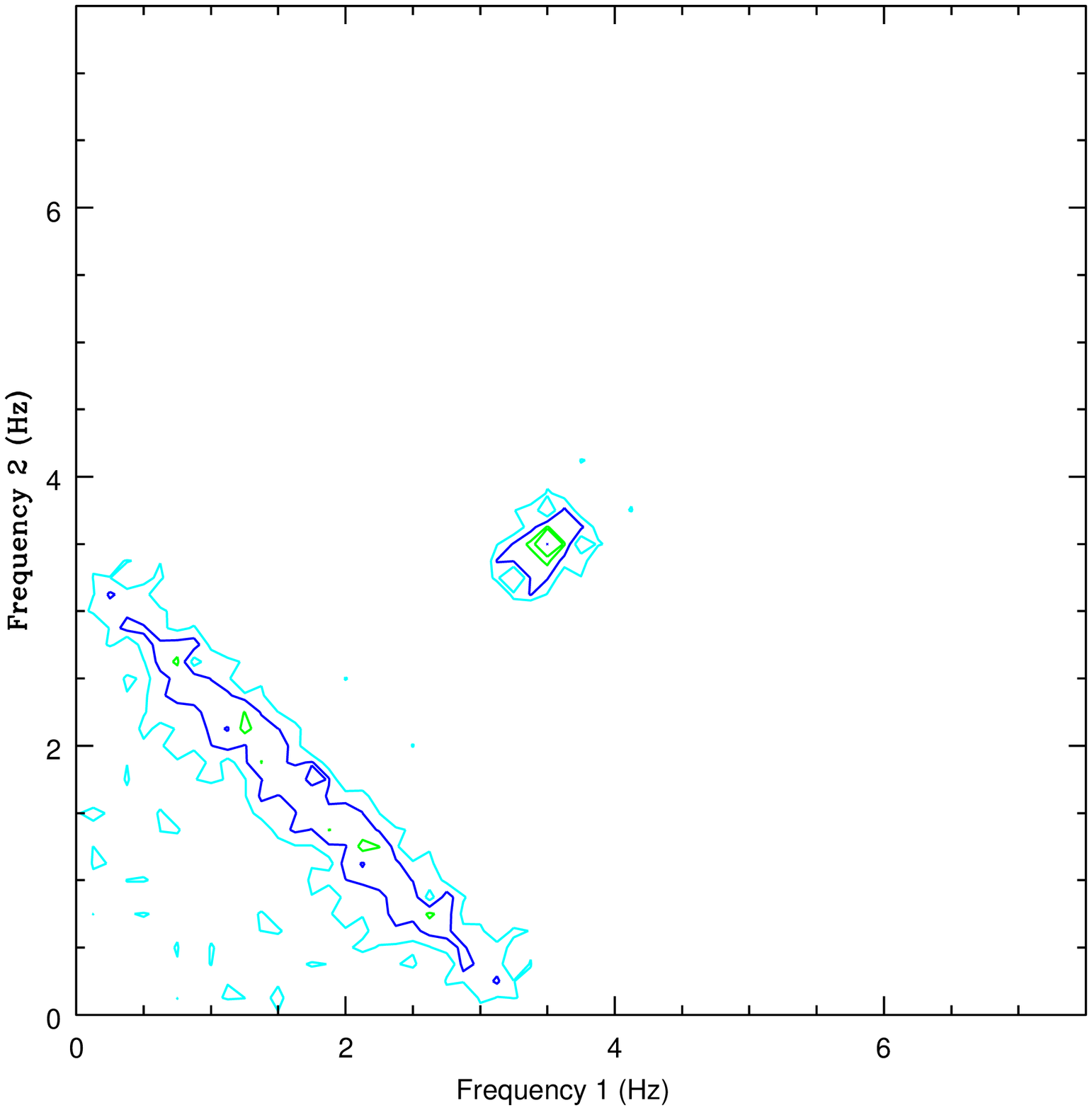,width=6cm}
\caption{The bicoherence plots from the observations.  (a) the data
for observation 10408-01-25-00, the prototypical ``web'' source. The
colour scheme is given for different values of the ${\rm log}_{10}
b^2$ as follows: red: -1.0; green: -1.25; dark blue: -1.5; light blue:
-1.75 (b) the data for observation 20402-01-15-00, the prototyical
``cross'' source. The colour scheme is given for different values of
the ${\rm log}_{10} b^2$ as follows: red: -1.0; green: -1.25; dark
blue: -1.5; light blue: -1.75 (c) the data for observation
30184-01-01-000, the prototypical ``hypotenuse'' source.  The colour
scheme is given for different values of the ${\rm log}_{10} b^2$ as
follows: red: -1.0; green: -1.25; dark blue: -1.5; light blue: -1.75.
The equality of the values of the bicoherence for reflections about
$x=y$ is trivial.}
\label{realbico}
\end{center}
\end{figure}

\begin{figure}
\begin{center}
\renewcommand{\epsfsize}[2]{0.7#1}
\epsfig{file=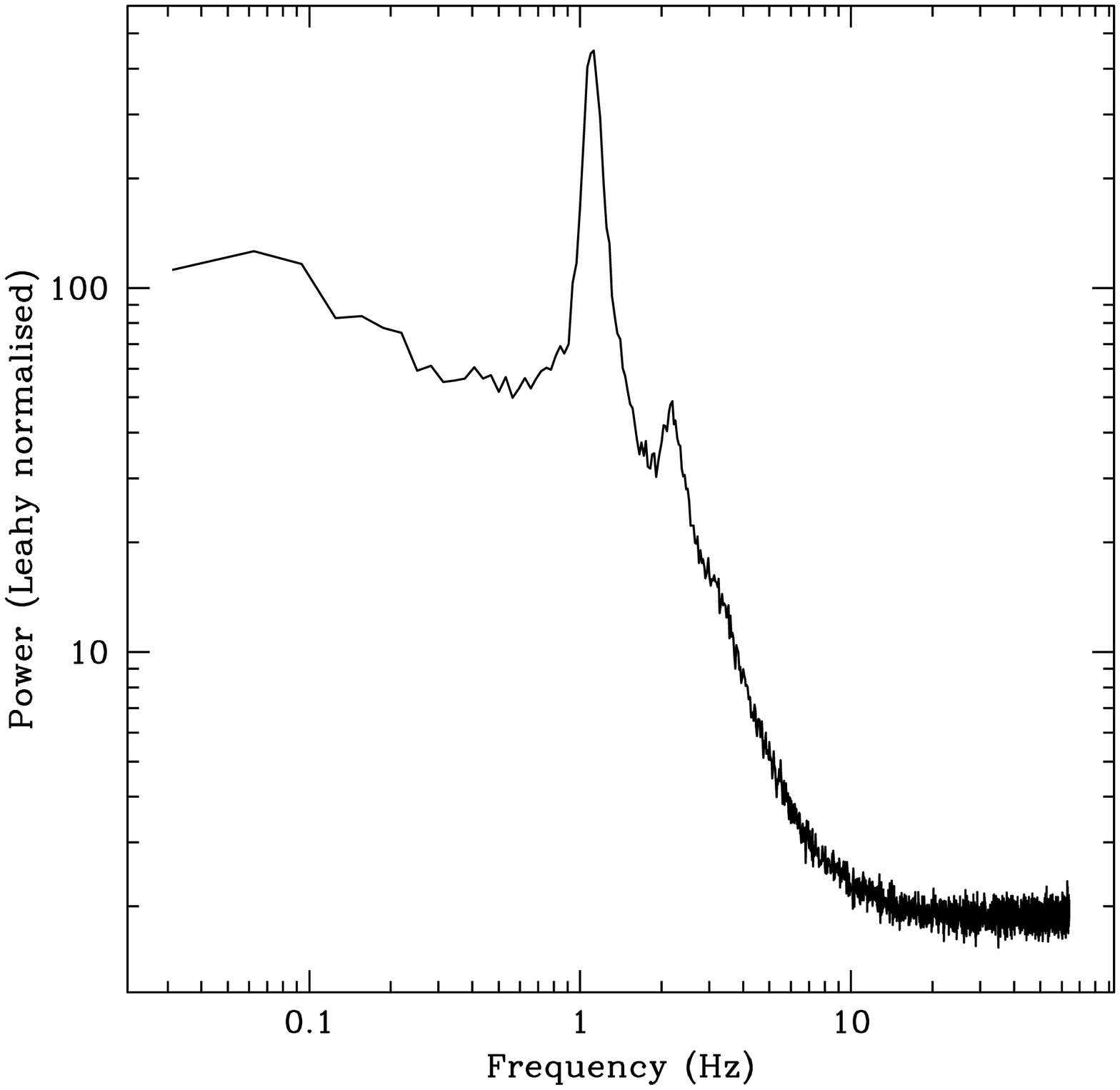,width=6cm}
\epsfig{file=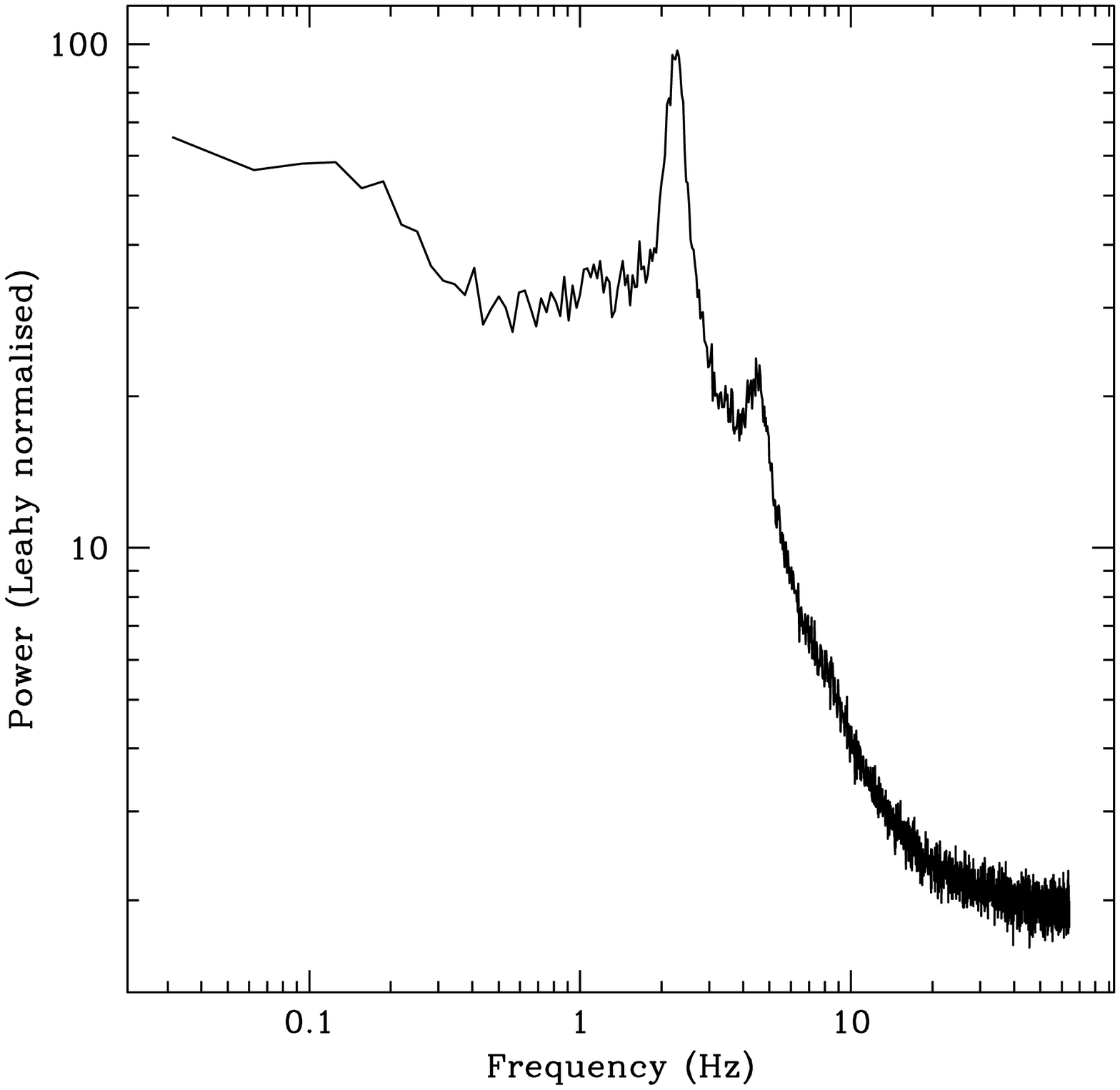,width=6cm}
\epsfig{file=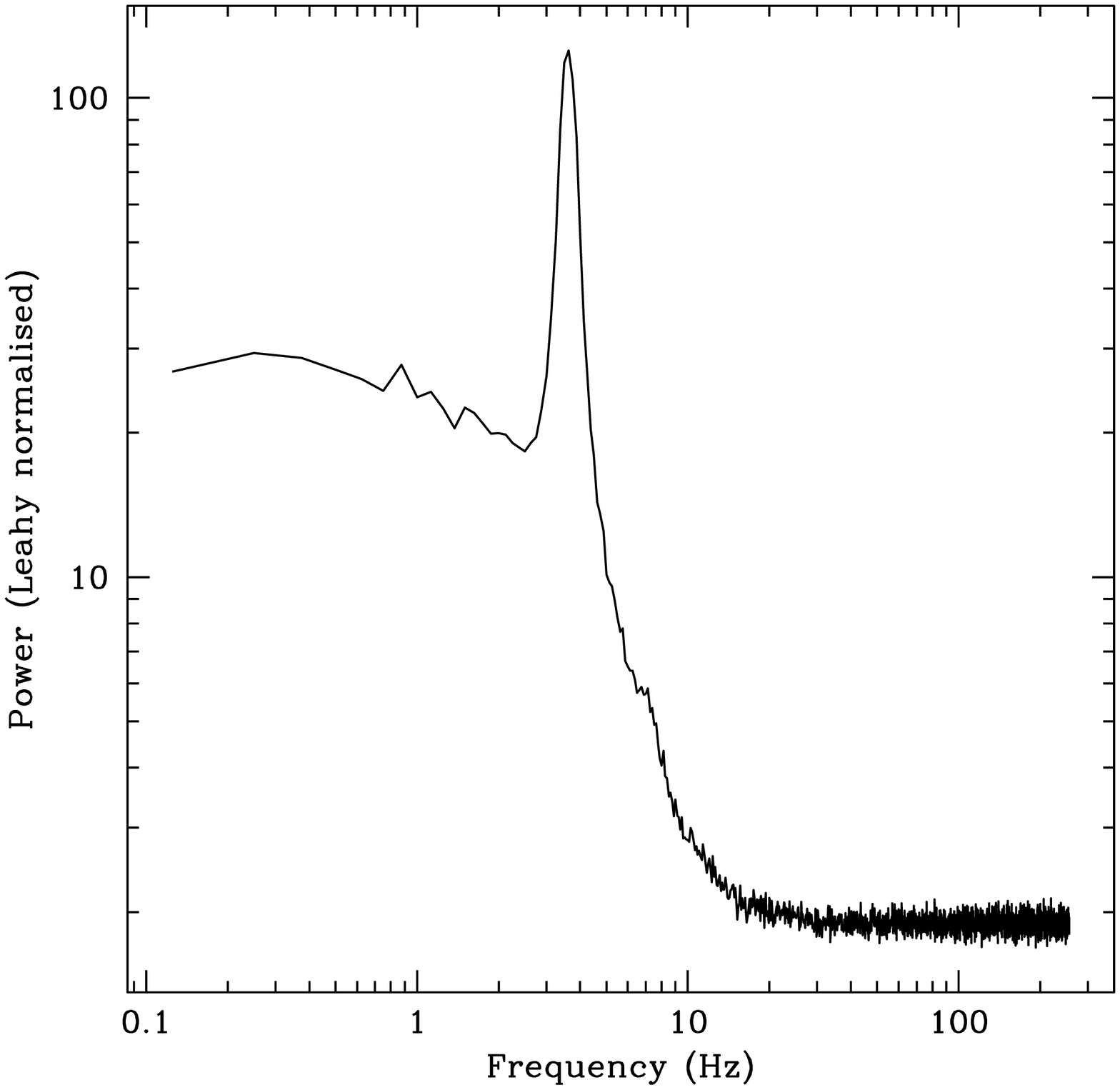,width=6cm}
\caption{The power spectra for the real data: (a) observation
10408-01-25-00 (b) observation 20402-01-15-00 (c) observation
30184-01-01-000.}
\label{realffts}
\end{center}
\end{figure}

We have computed the bicoherence for many observations of
GRS~1915+105.  For the sake of brevity, we present only three
representative results, aiming to present three observations with very
similar power spectra, but qualitatively different bicoherences.  We
find three phenomenological patterns of variability here, which we
described as ``web'', ``cross'', and ``hypotenuse''.  The bicoherences
from actual data are plotted in figure \ref{realbico}, and the
corresponding power spectra are plotted in figure \ref{realffts}.

\subsection{The ``cross'' pattern}
When the source is in the ``radio quiet'' state C at low count rates,
the QPO frequency is low (less than 2 Hz).  Here, the bicoherence
shows a ``cross'' behavior (see the results from ObsID 20402-01-15-00
plotted in figure \ref{realbico}b for the bicoherence and figure
\ref{realffts}b for the corresponding power spectrum), with large
bicoherence for frequency pairs where one frequency is the QPO
frequency and the other can take on any value.  Here, in contrast to
the web pattern, there is large bicoherence even for noise frequencies
less than twice $f_{QPO}$.  Additional power is seen at the harmonics
and for the harmonics interacting with the noise.

It is interesting to consider whether the cross pattern might be
indicating the effects of very broad wings to the QPO, with no actual
coupling between the QPO and the noise, but just coupling between the
QPO and its harmonics.  This is unlikely to be the case.  Firstly, the
bicoherence features are asymmetric in frequency space, while the QPOs
are well-fit by Lorentzians, indicating that they are not asymmetric
in frequency space.  In principle, since the bicoherence is normalised
by dividing by quantities related to the power spectrum, changes in
the strength of the noise component, which would dilute the coupled
variability, could be expected.  In this case, though, there is both
more power in the noise component of the power spectrum and stronger
bicoherence at frequencies below the QPO frequency than above it.
Secondly, the harmonics tend to follow either roughly circular
patterns for their bicoherence, if the quasi-periodicity is caused by
phase disconnections in an otherwise periodic signal, or $f_1=f_2$
patterns, if the quasi-periodicity is caused by the presence of many
real frequencies due to changes in frequency over time or the
superposition of multiple frequencies as might happen if the
variability is due to orbital motions over a range of radii (see
Maccarone \& Schnittman 2005), and what is observed is different from
both.

\subsection{The ``hypotenuse'' pattern}
At higher count rates of the ``radio quiet'' state C, the bicoherence
pattern changes dramatically (see the results from ObsID
30184-01-01-000 plotted in figure \ref{realbico}c for the bicoherence
and figure \ref{realffts}c for the corresponding power spectrum) .  A
high bicoherence is seen primarily where the two noise frequencies add
up to the QPO frequency -- that is, the regions of large bicoherence
make a diagonal line.  We refer to this as the ``hypotenuse'' pattern
because the region of strong bicoherence forms the hypotenuse of a
triangle with its other two sides being the axes of the plot.  The
``hypotenuse'' pattern typically also shows a roughly circular spot of
high bicoherence with $f_1$=$f_2$=$f_{QPO}$, since there is power both
at the fundamental and at the first harmonic and these two variability
components are coupled.  A few other observations at similar count
rates and with similar QPO frequencies in the non-plateau $\chi$ state
have been examined, and show the same qualitative features in their
bicoherences.  We also note that the diagonal elongation in the
bicoherence plot from lower left to upper right for $f_1=f_2=f_{QPO}$
is likely to be related to the frequency drift during the observation
that is mentioned above.

\subsection{The ``web'' pattern}
A hybrid class of bicoherence plots can be seen for some of the
observations, where features resembling those seen in the ``cross''
and of the ``hypotenuse'' patterns can be seen together.  When the
source is in the radio plateau state, as in ObsID 10408-01-25-00, the
bicoherence shows a rather strong signal for two combinations (see
figure \ref{realbico}a and figure \ref{realffts}a for the
corresponding power spectrum) -- a diagonal line from upper left to
lower right, with $f_1+f_2=f_{QPO}$, for all $0 < f_1,f_2 < f_{QPO}$.
Additionally, one vertical/horizontal streak is seen with one of the
frequencies equal to the QPO frequency, and the other a frequency
greater than the twice QPO frequency.  No significant
vertical/horizontal signal is seen for the case that $f_1< 2 f_{QPO}$.
The value of the bicoherence tapers off for frequencies where the
power in the noise component becomes small.  We refer to this pattern
of behavior as the ``web'' pattern.  A few other observations at
similar count rates in the plateau state have been examined, and show
the same qualitative features in their bicoherences.

\subsection{A few brief remarks on other observations}
In some observations not plotted in this paper, with the highest count
rates seen from GRS~1915+105 in its $\chi$ class, the bicoherence
seems to tend towards zero entirely.  However, at the highest count
rates, the QPO frequency can also change substantially on the
timescale of a single RXTE observation.  The bicoherence is not well
suited to dealing with data sets which are non-stationary in this
manner.  This problem may exist even in some of the lower count rate
data sets, but at a much lower level.  We note that in a few of the
observations, the bicoherence is extended along the line $f_1=f_2$
around the location where the interactions are between the QPO and its
first harmonic -- this effect can be seen showing up weakly in Figure
1a.  It was shown in Maccarone \& Schnittman (2005) that this is
characteristic of a QPO which is broadened due to the existence of
power at many frequencies, rather than due to phase disconnections.

\section{Discussion}
The overall values of the bicoherence in these data can also give us
some clues as to the types of processes that might be producing the
phase correlations.  The bicoherence relates the instantaneous
response of the driven oscillation to changes in the driving modes.
As stated above, Greb \& Rusbridge (1988) have shown that the maximum
value of the squared bicoherence is $\delta$$\omega$/$\Sigma$,
where$\delta$$\omega$ is the frequency resolution in the observation
(or the width of a QPO, if the QPO is resolved), and $\Sigma$ is the
frequency width of the driving spectrum.  Therefore, if a QPO is
driven by interactions of all noise frequencies less than the QPO
frequency, then the largest possible value of the squared bicoherence
is $Q^{-1}$, where $Q$ is the quality factor of the QPO.

Greb \& Rusbridge (1988) also consider the effects of having a
relatively narrow resonance.  In this case, the squared bicoherence is
reduced by a factor of the ratio of the coherence time in the driven
mode to that in the driving mode.  Thus, for resonant modes, which
remain coherent for long periods of time relative to the timescale on
which the driving signal remains coherent, the bicoherence will be
substantially reduced.  This can be grasped intuitively in a
qualitative sense -- a strong resonance will have an amplitude related
to the integral of the power dumped into it over a long timescale, and
hence not strongly correlated with the instantaneous driving power.

In our observations, the $Q$ values of the QPOs are typically $\sim3-10$,
and the QPOs are overresolved in frequency space by factors of about
10-20.  This yields expected maxima for the squared bicoherence of
$\sim$ 0.05 (with the exact value depending both on which observation
is considered, and whether multiple noise components interact
nonlinearly to produce the QPO, or the QPO interacts nonlinearly with
one noise component to produce another noise component).  Larger
values are possible for the harmonics of the QPOs, where the width of
the driving spectrum is smaller than the width of the noise component
in frequency space.

If we assume that the perturbations which are coupled to one another
give an X-ray count rate which is linearly proportional to the
amplitude of the perturbation, we can then use the magnitude of the
bicoherence to gain some insight into whether the interactions can be
through a narrow resonance at the QPO frequency.  Our findings that
the maximum values of the squared bicoherence for the cases where the
QPO and the noise are interacting are $\sim 10^{-1.5}$ would then
imply that the coupling between the noise and QPO is relatively close
to its maximal value.  This places some immediate constraints on the
classes of models which can be considered -- models must have a large
fraction of the power coming from an emission region which behaves as
a single ``system'', and, if the interactions between noise components
produce a QPO through a resonant interaction, it must be a highly
damped resonance.

\subsection{Simulated light curve analysis}

We now consider several different mathematical forms for light curves
which correspond, at least approximately, to physical scenarios for
producing QPOs and noise components which are correlated in some
manner.  We show that these different models can produce relatively
similar power spectra while producing bicoherence diagrams which are
qualitatively quite different from one another.  Two of the patterns
of behaviour found -- the ``cross'' and the ``hypotenuse'' -- are
reproduced reasonably well by relatively straightforward models.  The
third, the ``web'' is only approximately reproduced here with a
relatively simple model.

\subsubsection{Bicoherence in terms of reservoir models}

Where the bicoherence's value is large, the variability must be
coupled on the three timescales corresponding to the three frequencies
included for the computation of the bicoherence.  A natural way to
couple variability on different timescales is with a reservoir model,
where there are multiple variability components that drain the same
reservoir.  In this way, the different components ``compete'' for power --
when the noise component becomes stronger, there will be less energy
in the reservoir available for the QPO and vice versa, leading to a
phase coupling.  An analogous idea has been put forth to explain the
coupling between optical variability and X-ray variability in XTE
J1118+480 (Malzac, Merloni \& Fabian 2004).

To determine what the output light curve will look like for the
simulations, we define, in each case, two power spectra to use as
outputs.  One, the noise component, is a broad Lorentzian with a peak
frequency of zero.  The other, the quasi-periodic oscillation, is a
much narrower Lorentzian with a peak at a higher frequency (we use
$Q=30$ here, where $Q$ is the quality factor of the Lorentzian).  The
time series for these components are made using the method of Timmer
\& K\"onig (1995).  This method generates simulated times series from
a power spectrum under the assumption that the process is Gaussian --
thus we can be assured that any non-linearity we detect is because of
the physics we put in after generating the initial random time series.

Then, a reservoir is defined.  Energy is injected into the reservoir
either at a constant rate, or at a purely random rate (i.e. a random
number uniformly distributed between zero and one times a
normalisation).  Energy is drawn from the reservoir by each of the two
components.  The flux of a component is taken to be the value of its
time series multiplied by the size of the reservoir at that instant
times some normalization.  The flux is then subtracted from the
reservoir.  Specifically, we do the following:
\begin{equation}
R'(t)=R(t-\Delta{t})+Y_r(t),
\end{equation}
where $R'(t)$ is the size of the reservoir at time $t$, after the
energy injection has taken place, but before the draining of the
reservoir has taken place, $\Delta$$t$ is the time step,
$R(t-\Delta{t})$ is the size of the energy reservoir at the end of the
previous time step, and $Y_r(t)$ is the injection rate into the
reservoir per time step;
\begin{equation}
Y_{QPO}(t)=y_{QPO}(t)\times R'(t)\times k_{QPO},
\end{equation}
where $Y_{QPO}$ is the output flux from the QPO component, $y_{QPO}$
is the time series of the QPO produced by the Timmer \& K\"onig
method, and $k_{QPO}$ is a normalization constant; 
\begin{equation}
Y_{N}(t)=y_{N}(t)\times R'(t)\times k_{N},
\end{equation}
where $Y_{N}$ is the output flux from the noise component, $y_{N}$
is the time series of the noise produced by the Timmer \& K\"onig
method, and $k_{N}$ is a normalization constant; and, finally,
\begin{equation}
R(t)=R'(t)-Y_{QPO}(t-\Delta{t})-Y_{N}(t-\Delta{t}).
\end{equation}

Two general cases are considered, and we search a range of parameter
space for each case.  The first is where the output ``light curve'' is
the sum of the fluxes of the two components.  This might correspond to
a case where accretion power is dissipated either in some oscillating
region which produces the QPO, or a non-resonant part of the accretion
disc, which produces the noise component.  The second is where the
output light curve is only the QPO's flux.  This might correspond to
the case where the broad Lorentzian represents a jet which extracts
power from the system, but does not emit in the X-rays (see
e.g. Malzac et al. 2004).

If the reservoir is drained by a component which does not contribute
to the X-ray light curve, then the size of the reservoir in ``view''
of the X-ray light curve, will effectively be some quantity minus the
integral of the power drained by the other component.  As a result,
the reservoir itself will have a signature of its modulation, so when
the output light curve is obtained by multiplying the reservoir size
by the time series for the QPO, the ``QPO'' component will now be
modulated on the timescale of the ``noise,'' and the noise power
spectrum will affect final power spectrum.  As we show below, the
qualitative properties of the bicoherence are largely the same
regardless of whether the two drains on the energy reservoir both
contribute to the output flux, or only one of these does.  Again, the
logic we follow here is quite similar to that in Malzac et al. (2004).
Given that, as we will show below, these models seem to match the data
for the plateau states more closely than the more radio quiet $\chi$
states, it would not be surprising if energy extraction by a jet were
an important factor in determining the properties of the light curve.

After the basic model equations are defined, a time series is produced
for the model.  The QPO and noise components are produced as time
series using the Timmer \& K\"onig method, with both components
assumed to be Lorentzians with parameter values set to values similar
to those seen in the data for various observations.  For a very small
number of time bins, the value of one of these time series will be
negative, in which case we set it to zero.  Alternatively, the energy
extracted in a time interval may be larger than the size of the
reservoir, in which case we set the power such that the whole
reservoir is drained in that time step.  

When the amplitude of variability is large, the reservoir's level
fluctuates strongly.  This leads to the production of harmonics in the
QPO (since the oscillations become non-sinusoidal when the reservoir
fluctuates in response to the oscillations themselves).  It also leads
to phase coupling of all types between the QPO and the noise.  This
produces a ``web-like'' pattern in the bicoherence; however, the
web-like pattern here is not identical to the one seen in the radio
plateau state.  In the real data, the ``cross-like'' structures begin
to manifest themselves only for noise frequencies larger than the
frequency of the first harmonic, while in these simulations, the cross
structures appear at all noise frequencies.  Furthermore, in some of
the simulations, there are diagonals from upper left to lower right
related to the cross structure which are strong features at all
frequencies adding up to at least one of the harmonics.  These are not
present in the real data.  The strong diagonals manifest themselves
when the total variability in the reservoir is dominated by the noise
component, rather than by the QPO.  The model bicoherence is shown in
Figure 3b to represent a case where the jet extracts power (i.e. so
that the extracted noise-component power does not lead directly to
observed emission), and in Figure 4b for the case where the noise
component is added back in to the model light curve.  Figure 5b shows
the simulated power spectrum corresponding to the bicoherence shown in
Figure 3b.

In the numerical calculations where the noise component is not added
to the final light curve, the QPO has a mean count rate of 8000, with
a root mean squared variability level of 1200, while the noise
component has a mean value of 20000, with an root mean squared
variability level of 8000.  We re-fill the reservoir by adding a
random number drawn from a uniform distribution between 0 and 8000.
The draining of the reservoir takes place by summing the two time
series, multiplying by the size of the reservoir, dividing 200000, and
subtracting that from the reservoir value.  

Where the noise component is added to the final light curve, a
slightly different procedure is adopted.  The QPO and noise components
are added together in the initial power spectrum made before the
inversion into a time series using the Timmer \& K\"onig method.  The
QPO is given a normalization in the power spectrum 200 times higher
than the noise component's.  The actual value in the final power
spectrum comes from converting this power spectrum into a time series
with mean value of 8000 and rms amplitude of 6000.  All other
procedures are as above.

At low variability amplitude, only the cross-like structures are seen
strongly, although a weak ``hypotenuse'' feature is still present when
the noise component is added to the output ``light curve.''  The
bicoherence is shown without the noise component added back in in
Figure 3a, and with it added back in in Figure 4a.  The power spectrum
corresponding to the low variability amplitude model is shown in
Figure 5a.  The simulated bicoherence plots in this case look quite a
bit like the bicoherence plot seen from observation 20402-01-15-00
(compare the observational data in Figure \ref{realbico}b with the
simulations in Figure \ref{modelbico}a).

The procedure for converting the model into a simulated light curve is
the same as above, except for some changes in values of parameters.
For the case where the noise component is not seen in the output
simulated light curve, the noise component in this case has an rms
amplitude of 1000 instead of 8000, with all other parameter values the
same.  Where the noise component is added to the final simulated light
curve, the rms amplitude of the simulated light curve is 2500 instead
of 6000.

\subsubsection{Bicoherence in terms of damped forced harmonic oscillators}

We have also considered the case of a damped, forced harmonic
oscillator.  This is conceptually similar to the mechanism suggested
by Psaltis \& Norman (2000) for producing both quasi-periodic
oscillations and noise components from a single perturbation spectrum
-- the quasi-periodic oscillation appears at the resonant frequency of
the system, while a white noise input spectrum is turned into a red
noise component by the fact that damped oscillators act as low-pass
filters.  However, the response of the system was considered to be
linear in that work, meaning that non-linear features like non-zero
bicoherence can not be explained without some modifications.  When
modifying the system of equations, to allow for a non-linear restoring
force, we find that the resulting pattern of bicoherence that results
from this scenario is relatively similar to the ``hypotenuse''
pattern, although we do have trouble reproducing a strong bicoherence
with a realistic power spectrum.  An alternative physical mechanism
for producing such a quasi-periodic oscillation would be to have
variations in the accretion rate excite inertial-acoustic modes in the
inner accretion disc (Chen \& Taam 1992;1995).

Quantitatively, this scenario is modelled in the same way as a spring
which does not follow Hooke's Law (i.e. $F=-kx$), but rather has an
asymmetric, higher order restoring force, $F$ which is given by
$F=-k_1 x + k_2 x^2$, where $x$ is the displacement, and $k_1$ and
$k_2$ are constants parameterizing the strength of the restoring
force.  If $k_2=0$, this does reduce to Hooke's Law, and the lack of a
non-linear term leads to a lack of observable features in the
bicoherence plot.

The full equations of motion for this ``oscillator'' are:
\begin{equation}
F(t)=-k_1 x(t) + k_2 x^2(t) -\gamma v(t) + Y_{WN}(t),
\end{equation}
where $\gamma$ is the damping coefficient and $Y_{WN}$ is the
magnitude of the driving force, which is assumed to be a white noise
process;
\begin{equation}
v(t)=v(t-\Delta{t})+\Delta{t}\times F(t)/m,
\end{equation}
where $m$ is the mass of the oscillator, which we set to unity for
these calculations, and $\Delta{t}$ is 0.001.  The calculated
simulated light curve is the time series $x(t)$, and has 1048576
points.  The bicoherence is computed from 256 chunks of 4096
points.

\begin{figure}
\begin{center}
\renewcommand{\epsfsize}[2]{0.7#1}
\epsfig{file=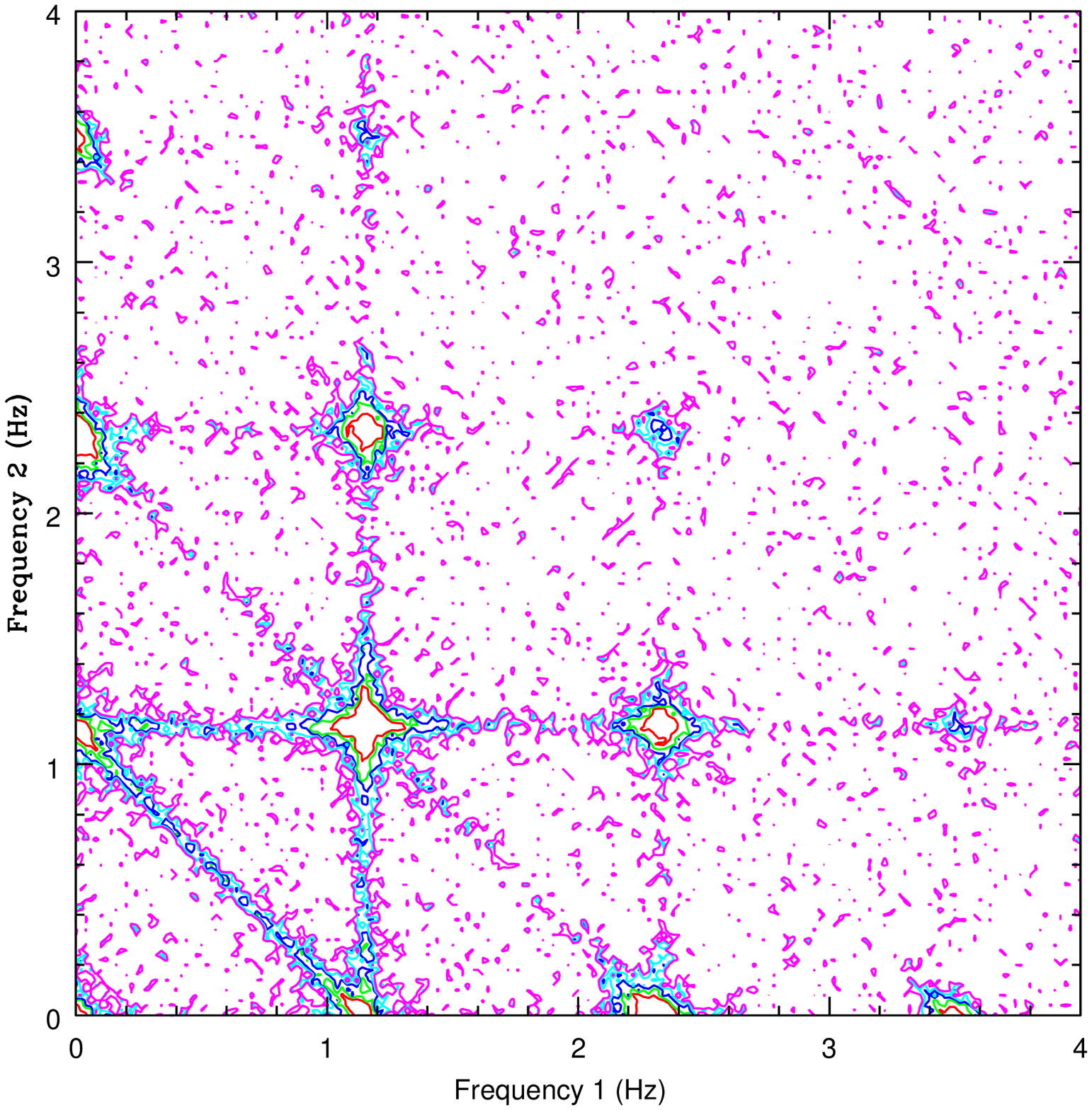,width=6cm}
\epsfig{file=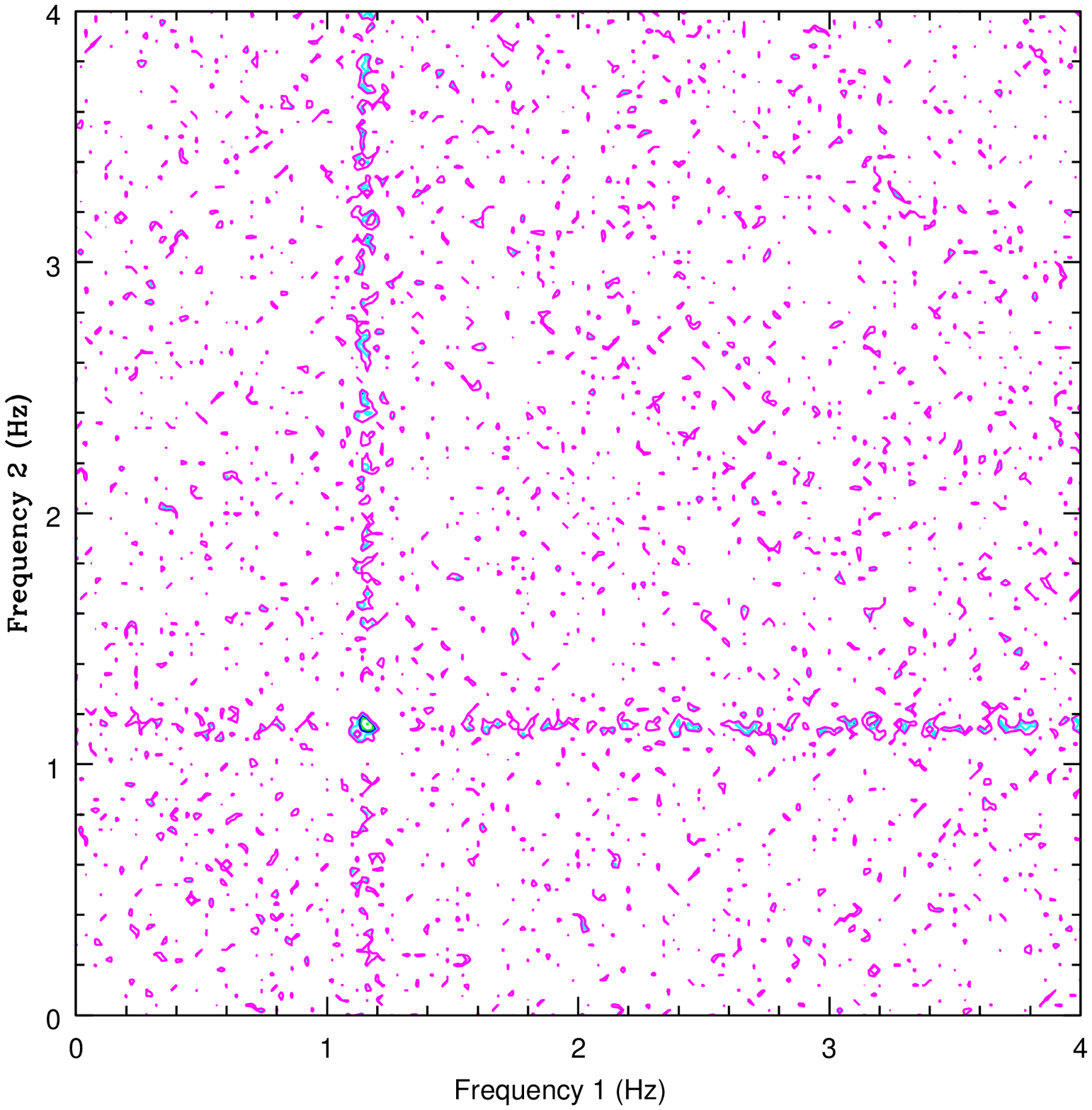,width=6cm}
\epsfig{file=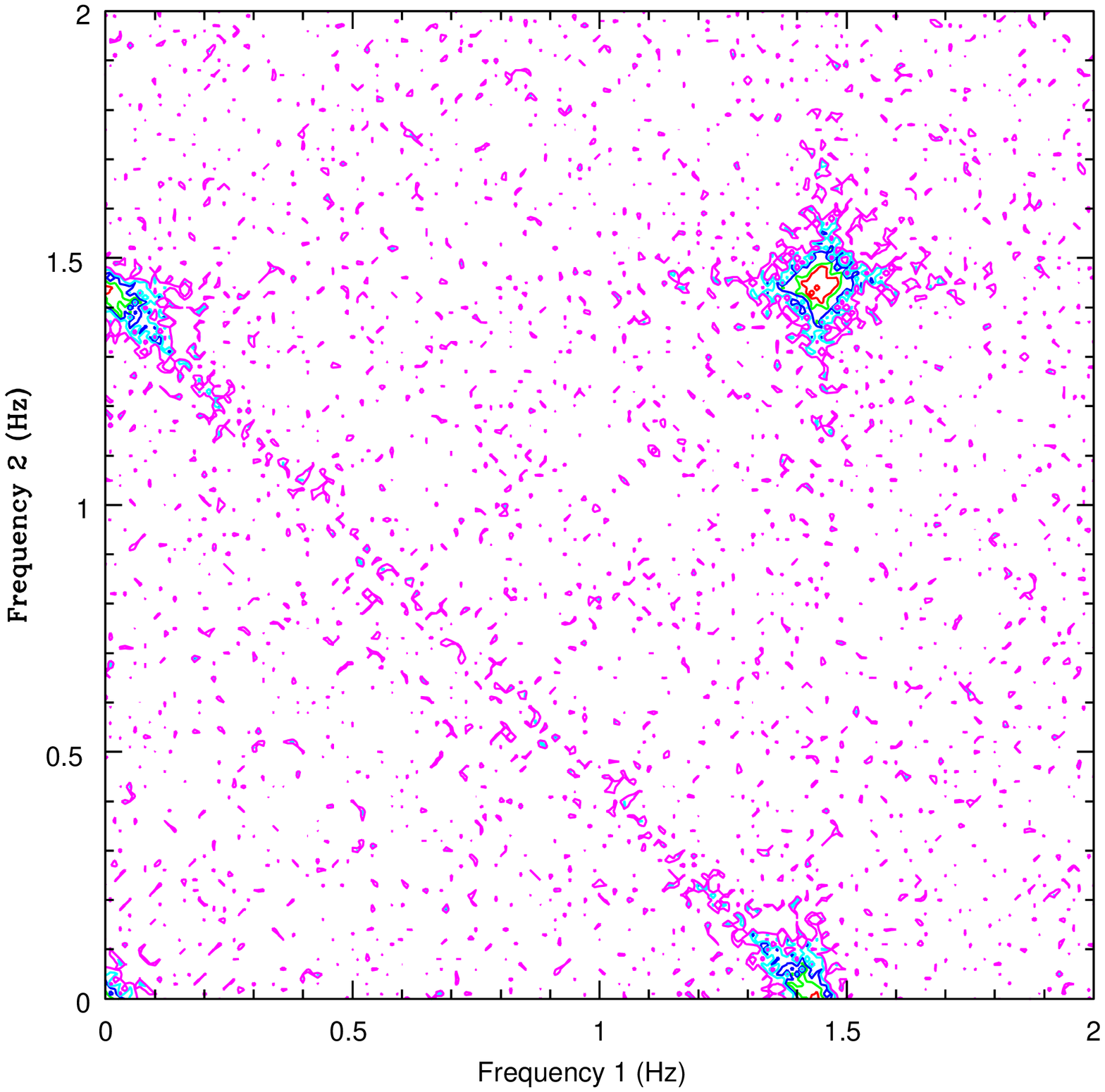,width=6cm}
\caption{The model bicoherence plots.  For all plots, the colour
schemes are as for the real data, except that for the model plots,
there is an additional contour, in purple, at ${\rm log}_{10}
b^2=-2.0$.  The plots are: (a) the case of ``jet'' extraction with a
high variability amplitude (b) the case of ``jet'' extraction with a
low variability amplitude (c) damped forced oscillations.  The
equality of the values of the bicoherence for reflections about $x=y$
is trivial.}
\label{modelbico}
\end{center}
\end{figure}

For the first calculation, we set the parameter values to
$k_1=50000.0 , k_2=0.2,$ and $\gamma=7.0$.  We show the plot of
simulated data in \ref{modelbico}c.  We note there is one substantial
difference between the model data and the real data for
30184-01-01-000, which is that the model data has stronger bicoherence
along the ``hypotenuse'' when the two frequencies are most different
from one another, while the real data shows the opposite trend.  This
trend makes the model values more seriously deviant from the real data
for the strongest damping of the motion of the oscillator.

When the damping factor is very low, then the resulting bicoherence
plots can look quite similar to the ``web'' pattern.  In figure
\ref{web_df}, we show the power spectrum and bicoherence plot for a
low damping case.  The calculations are done in the same manner as the
ones described in the previous paragraph, except with $k_2$=0.25 and
$\gamma=0.6$.  For values of $\gamma$ intermediate between 0.6 and
7.0, the ``hypotenuse'' can become much broader without creating such
strong additional harmonics in the power spectrum.  This simulation,
like the one whose power spectrum and bicoherence are shown in Figures
4\ref{modelffts}b and \ref{modelbico}b, respectively, matches many of
the properties of the data, but not all; however, here, the problem is
primarily that the model shows too strong a bicoherence in streaks
where there is only noise, rather than power in the QPO or its
harmonics.

\begin{figure}
\begin{center}
\renewcommand{\epsfsize}[2]{0.7#1}
\epsfig{file=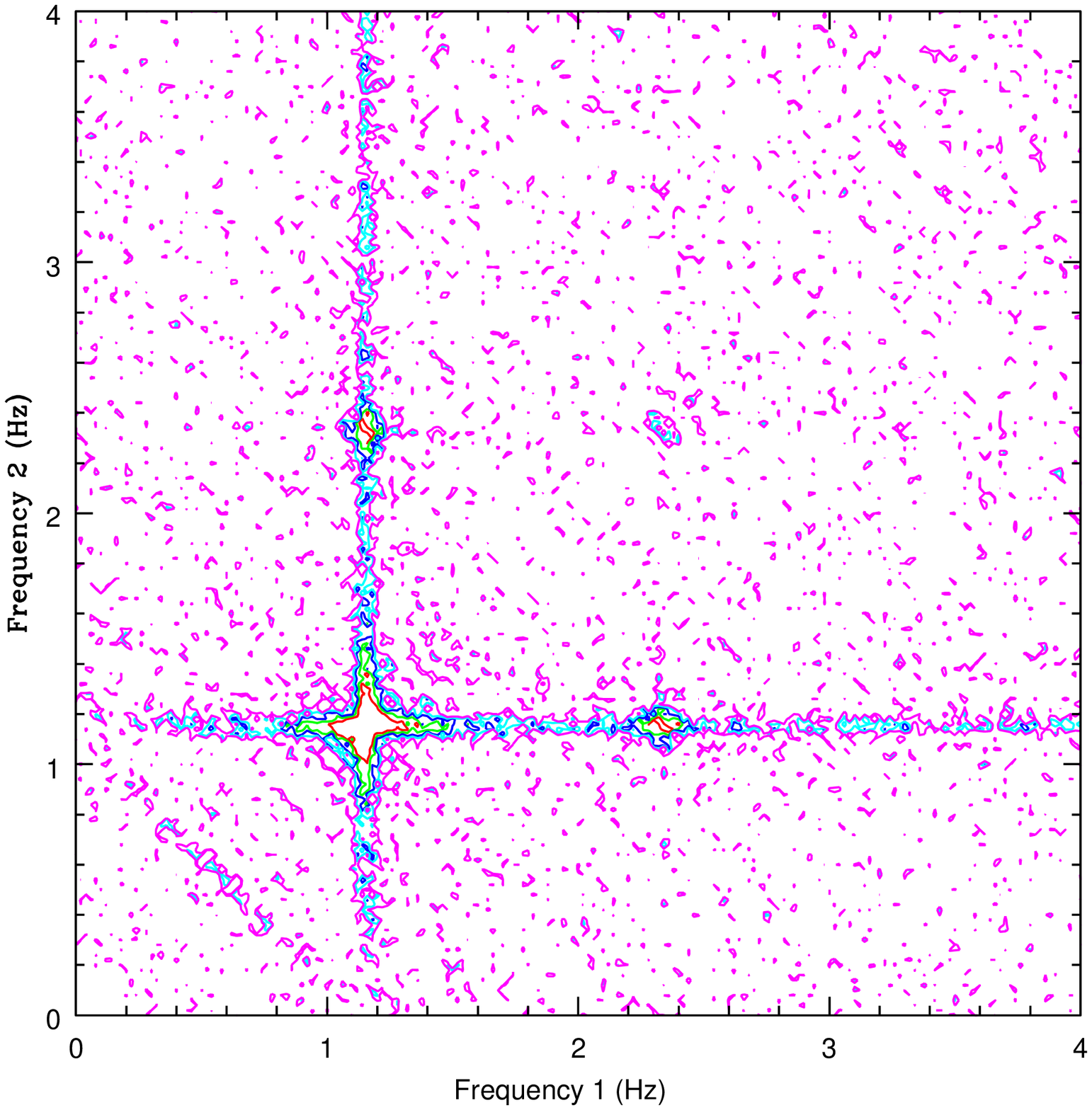,width=6cm}
\epsfig{file=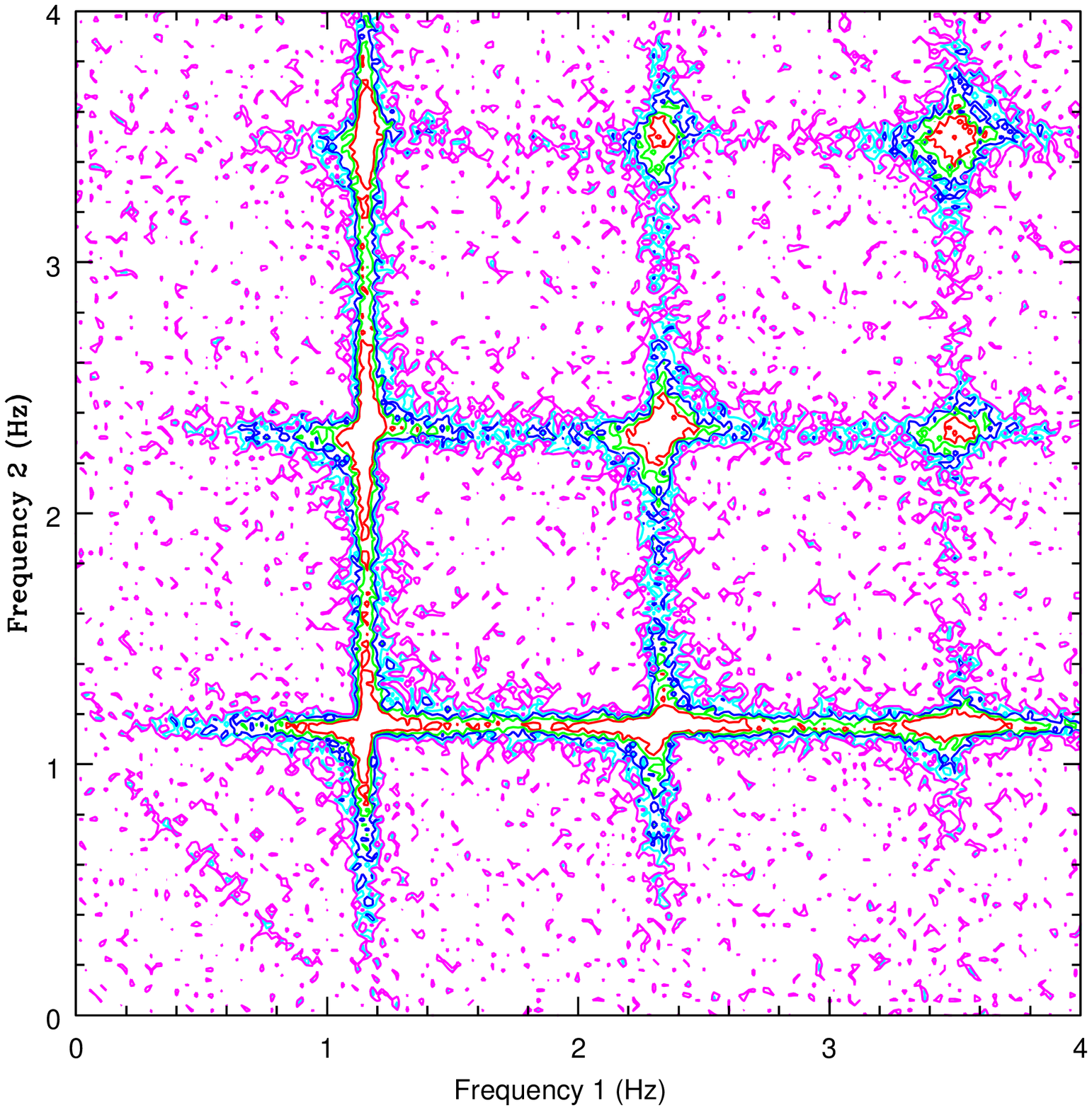,width=6cm}
\caption{The model bicoherence plots for reservoir models where the
noise component contributes to the observed flux.  For all plots, the
colour schemes are as for the previous plot of simulated data.  The
plots are: (a) low variability amplitude (b) high variability
amplitude. The equality of the values of the bicoherence for reflections about
$x=y$ is trivial.}
\label{modelbico2}
\end{center}
\end{figure}

\begin{figure}
\begin{center}
\renewcommand{\epsfsize}[2]{0.7#1}
\epsfig{file=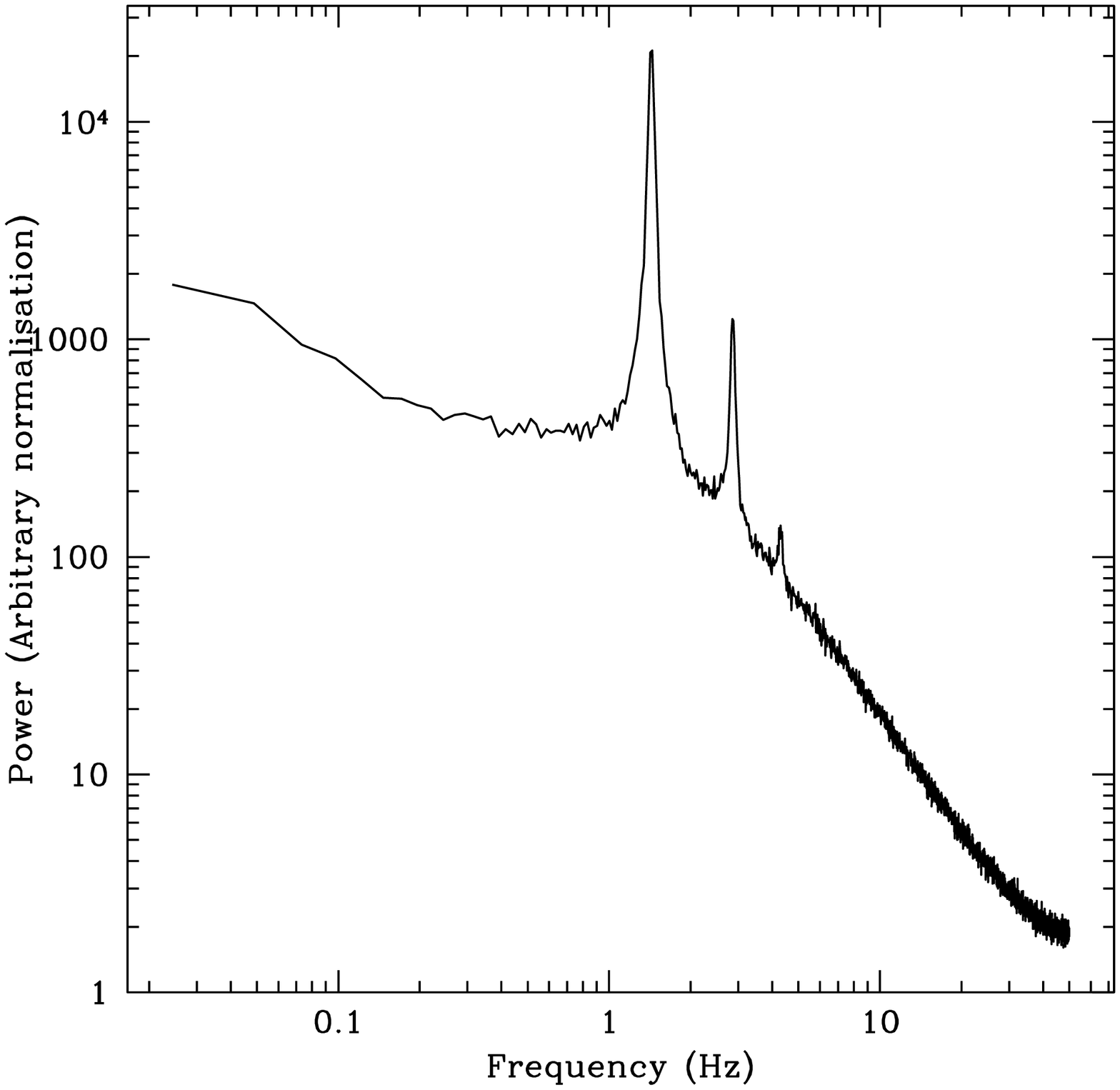,width=6cm}
\epsfig{file=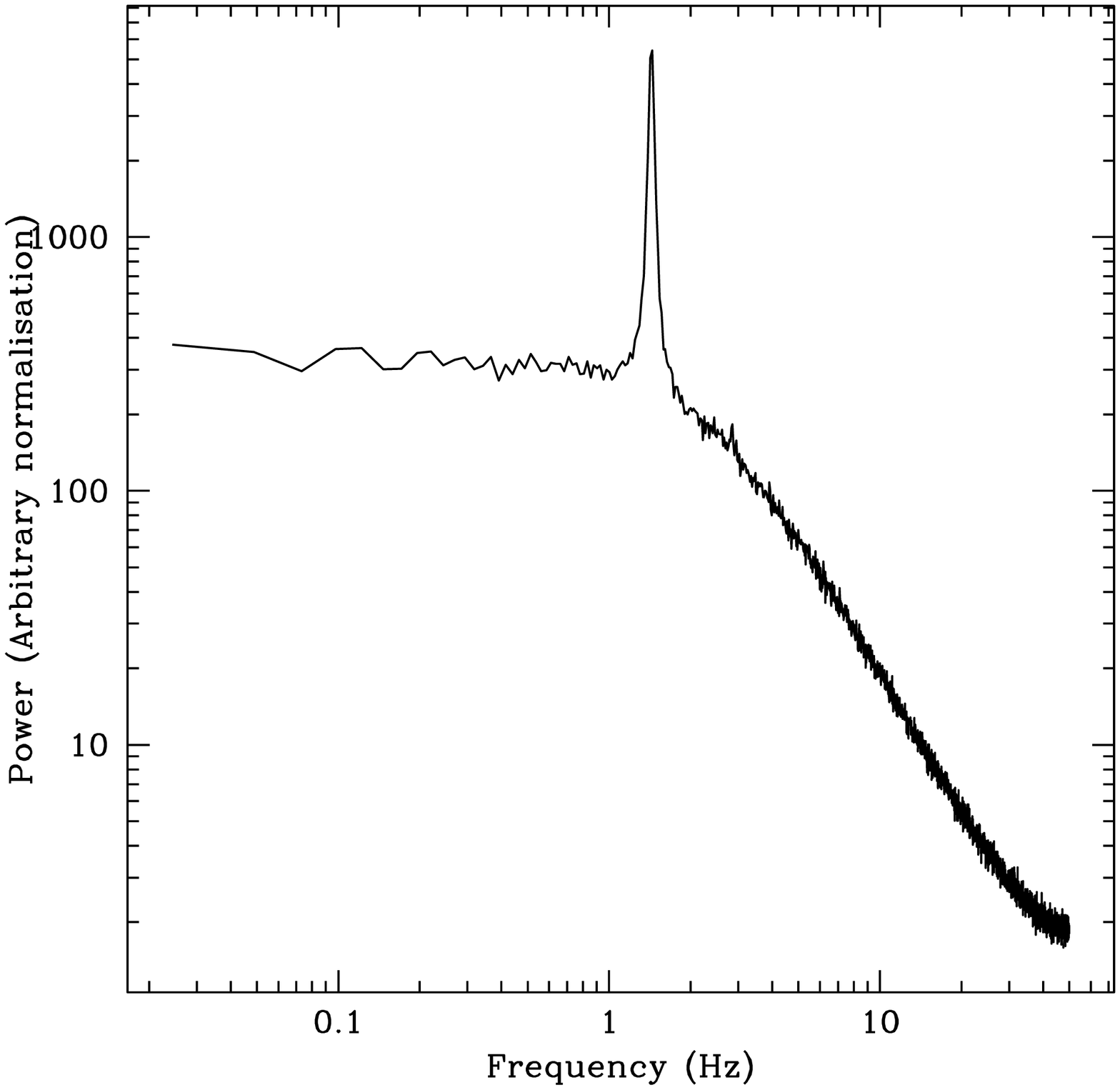,width=6cm}
\epsfig{file=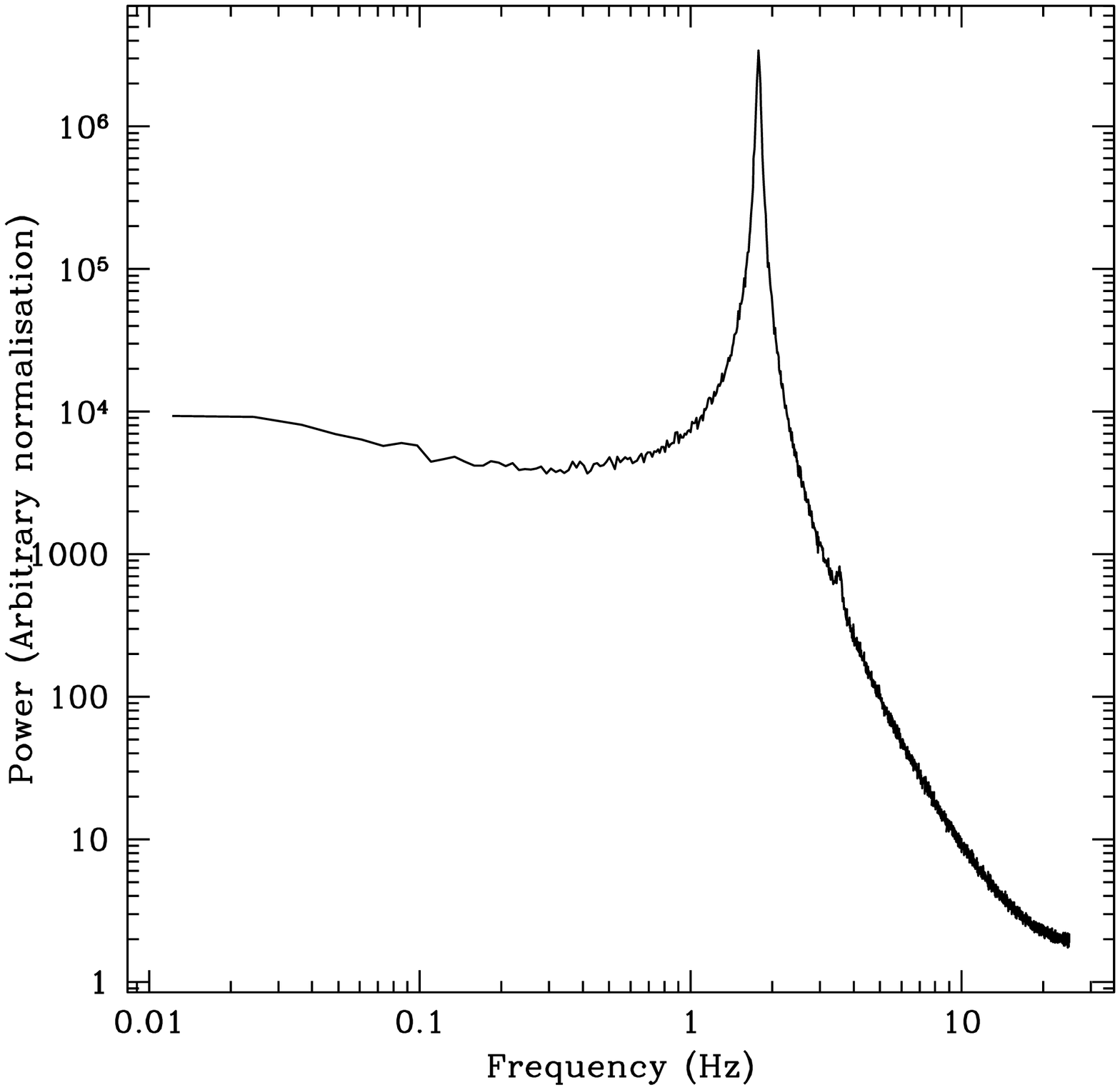,width=6cm}
\caption{The power spectra for the simulated data: (a) web-like (b)
cross-like (c) damped forced oscillations -- note that these are the
power spectra for the same time series, in the same order, as
presented in Figure 3.}
\label{modelffts}
\end{center}
\end{figure}

\begin{figure}
\begin{center}
\renewcommand{\epsfsize}[2]{0.7#1}
\epsfig{file=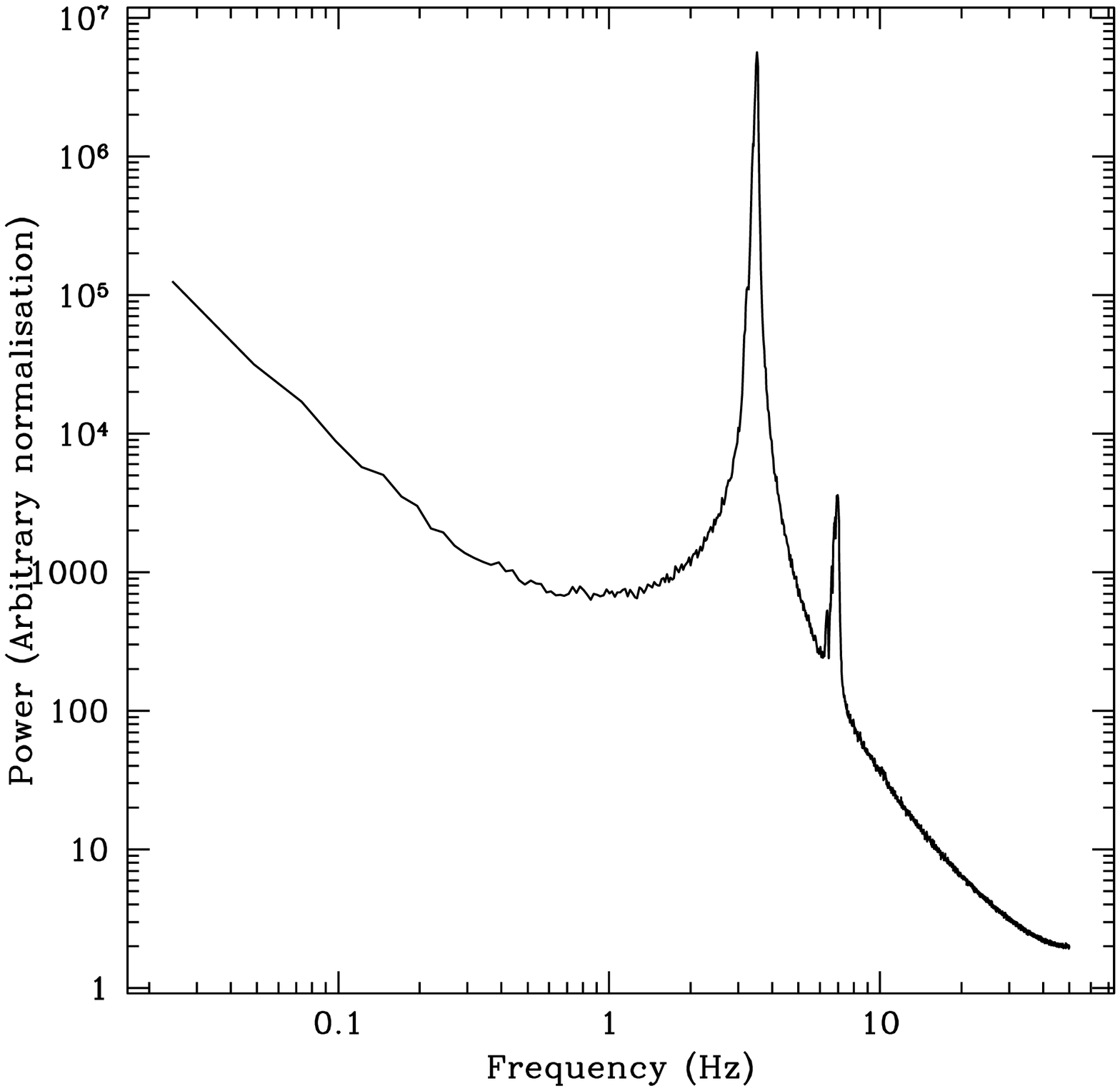,width=6cm}
\epsfig{file=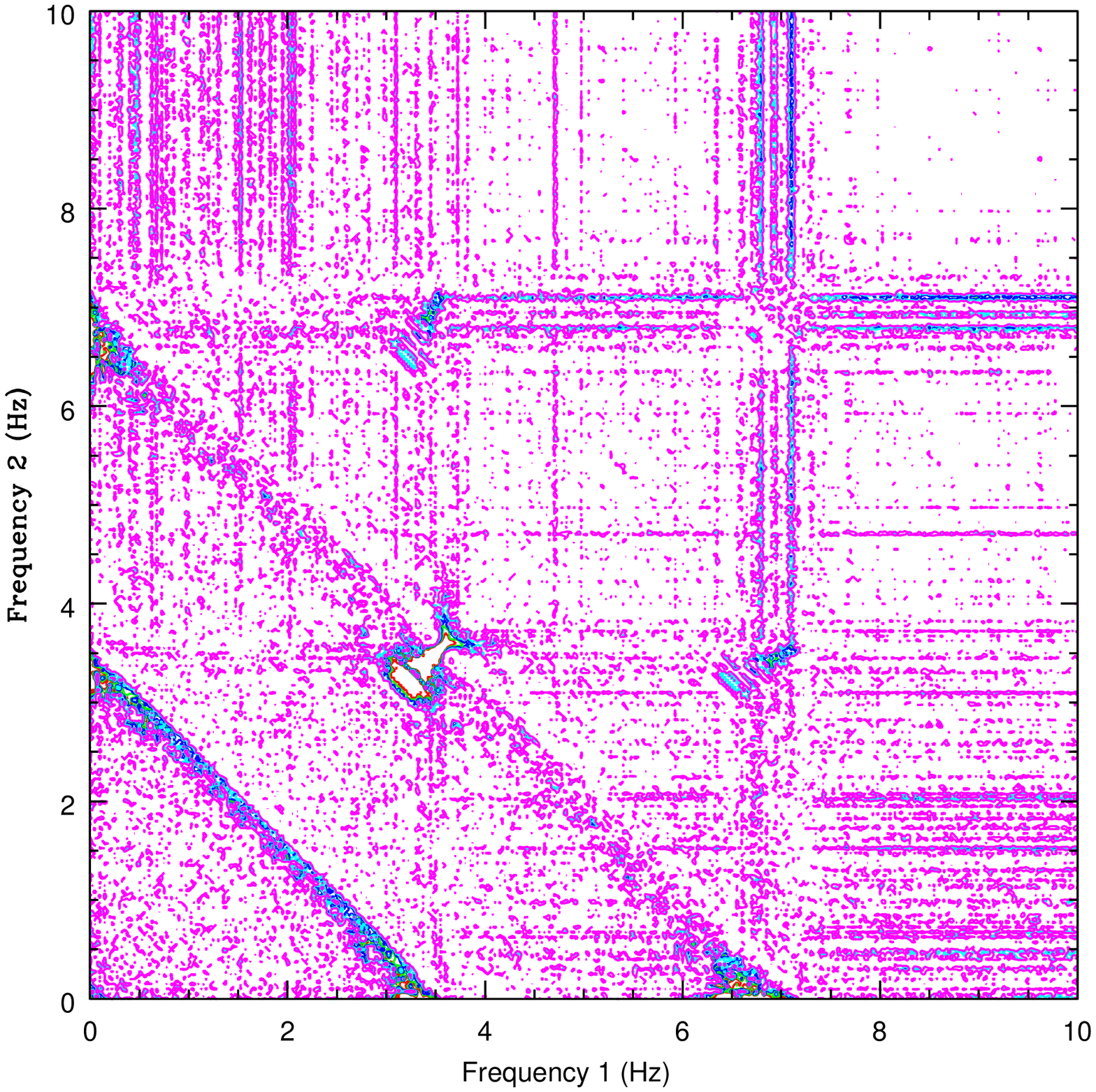,width=6cm}
\caption{(a) the power spectrum for the low-damping forced oscillator
(b) the bicoherence for the low damping forced oscillator, with the
same contour values as used in the other simulations' plots.  The
equality of the values of the bicoherence for reflections about $x=y$
is trivial.}
\label{web_df}
\end{center}
\end{figure}

\subsubsection{Caveats}
There are further steps which will be needed to make models which
match all the statistical properties of the observed light curves.
For example, in most cases, the observed flux distributions will
follow a log-normal distribution (Uttley et al. 2005) - we do not
attempt here to enforce this requirement.  We do note that it was
found by Uttley et al. (2005) that exponentiating the flux values in a
light curve can push a flux distribution towards a log-normal
distribution.  We have done this in a few cases, and have found that
the bicoherence plots change very little when the flux values are
exponentiated.  Uttley et al. (2005) have already shown that power
spectra are not typically affected strongly by this transformation.
Therefore, flux distributions and bicoherence plots are largely
independent tests of the variability's coupling, although the flux
distributions produced in these simulated data sets do not deviate
strongly from log-normal distributions in any event.  Furthermore, in
cases where there are strong QPOs with very low frequencies, there are
clear deviations of the flux distribution from being log-normal (Misra
et al. 2006).  We have additionally not ensured that the power spectra
of real observations are matched exactly by these simulations.  In
essence, since the input models do not have any radiative transfer in
them, we are assuming that the radiative efficiency of any
perturbation will be constant when we call the output time series
``simulated light curves.''  Deviations from this assumption are
likely to affect the power spectra, and, especially, the flux
distributions, far more seriously than they affect the morphology of
the bicoherence plots.  Such deviations are quite easy to imagine in
situations where there exists a resonance.

An additional potential problem exists with connecting the toy models
presented here to reality. Much of the phenomenology appears to be
similar between the different sub-types of $\chi$ classes.  It would
therefore be surprising if there were a qualitative change in the
mechanism for producing the QPOs.  It would thus be useful to identify
a means to unite the different mechanisms described here, or to
develop an entirely new physical model to explain all the observed
phenomenology in a more unified way, with just the change of one or
two parameters causing the broad range of phenomena seen in terms of
bicoherence patterns.

\section{Conclusions}
We have shown clearly that the variability in the noise components and
quasi-periodic oscillations of GRS~1915+105 are correlated with one
another.  The bicoherence provides a good discriminator among
variability models which produce similar power spectra through quite
different physical processes.  We have found that it is plausible that
in the ``plateau'' state of GRS 1915+105, the variability is caused by
a reservoir of energy being drained by a noise component (which could
be the radio jet) and a quasi-periodic component, while in the
brighter part of the $\chi$ state, the variability is consistent with
a white noise input spectrum driving a damped harmonic oscillator with
a non-linear restoring force.  While the models presented here are
almost certainly not unique solutions for what is occuring physically
in these systems, it is quite clear that the bicoherence will provide
excellent constraints on more physically motivated models for the
variability, and that we can definitely identify cases where the
properties of the power spectra are quite similar, but the properties
of the light curves are quite different.

\section{Acknowledgments}
We are grateful to Ron Elsner for sending us an unpublished draft
written by himself, R. Bussard and M. Weisskopf many years ago
discussing some of the fundamental properties of the bispectrum, and
to Patricia Ar\'evalo, Mike Nowak and Simon Vaughan for useful
discussions related to variability statistics, and to Marc Klein-Wolt,
Mariano M\'endez and Luciano Rezzolla for more general, but equally
interesting discussions, and to Rob Fender for helping to clarify some
of the nomenclature surrounding GRS~1915+105, and Tomaso Belloni for
useful comments on a draft of the paper.  We also thank the anonymous
referee for some useful suggestions which have hopefully improved the
clarity of the paper.  TJM also wishes to thank Dave Smith of the
Quantum and Functional Matter group in the Southampton School of
Physics and Astronomy for useful discussions.

\label{lastpage}
\end{document}